\documentclass[aps,prl,twocolumn,showpacs,longbibliography,superscriptaddress]{revtex4-1}

\usepackage{color}

\usepackage{graphicx}
\begin{document}
\newcommand{\ud}{{\mathrm d}}
\newcommand{\sech}{\mathrm{sech}}
%\preprint{}

\title{Avoided crossing and sub-Fourier sensitivity in driven quantum systems}

\author{David Cubero}
\email[]{dcubero@us.es}
\affiliation{Departamento de F\'{\i}sica Aplicada I, EUP, Universidad de Sevilla, Calle Virgen de \'Africa 7, 41011 Sevilla, Spain}

\author{ Gordon R.M. Robb}
\email[]{g.r.m.robb@strath.ac.uk}
\affiliation{SUPA and Department of Physics, University of Strathclyde, John Anderson Building, 107 Rottenrow, Glasgow, G4 0NG, UK}

\author{Ferruccio Renzoni}
\email[]{f.renzoni@ucl.ac.uk}
\affiliation{Department of Physics and Astronomy, University College London, Gower Street, London WC1E 6BT, United Kingdom}

%\author{David Cubero$^{1}$ and Ferruccio Renzoni$^{2}$}
%%%\email{$^{1}$ dcubero@us.es, $^{2}$ f.renzoni@ucl.ac.uk}
%\affiliation{$^1$Departamento de F\'{\i}sica Aplicada I, EPS, Universidad de Sevilla, Calle Virgen de \'Africa 7, 41011 Sevilla, Spain}
%\affiliation{$^2$Department of Physics \& Astronomy, University College London, Gower Street, London WC1E 6BT, UK}

\begin{abstract}
The response of a linear system to an external perturbation is governed by the Fourier limit, with  the inverse of the interaction time constituting a lower limit
for the system bandwidth. This does not hold for nonlinear systems, which can thus exhibit sub-Fourier behavior. The present work identifies 
a mechanism for sub-Fourier sensitivity in driven quantum systems, which relies on  avoided crossing between Floquet states. Features up to three orders of magnitude finer than the Fourier limit are presented.

\end{abstract}

\maketitle

Many processes in Physics can be traced back to the response of a system to a periodic perturbation of a finite 
duration  $T_s$. For a linear system its frequency response to the perturbation is governed by the Fourier limit: the system 
bandwidth $\Delta \omega$ cannot be smaller than the inverse of the interaction time, $2\pi/T_s$, the two quantities being equal in the absence of additional relaxation processes.

Nonlinear systems are not Fourier limited, and their response to an external perturbation may exhibit sub-Fourier features. 
It was in particular shown that for the kicked rotor system quantum interference leads to a sub-Fourier response 
\cite{garreau02,gardiner04,summy10}. Sub-Fourier behaviour was also observed in bi-harmonically driven classical systems \cite{cubero2013}. In  both cases, the underlying mechanism relies on the system's extreme sensitivity to the  nature of the driving, with periodic and quasiperiodic drivings leading to a completely different response.

In this work we identify a mechanism for sub-Fourier sensitivity in driven quantum systems, which relies on  avoided crossing between Floquet states. The quantum features identified here are found to be narrower than their classical counterparts, and results for features up to three orders of magnitude finer than the Fourier limit are presented.  These results are of general interest in terms of fundamental physics, as they allow a direct comparison with the classical counterpart, and thus provide a general framework to study the  role of quantum effects in the occurence of sub-Fourier dynamics in ac driven systems. Also they may find direct application in quantum control and sensing applications, where extreme sensitivity to external perturbations are of paramount importance.  Specifically, the ability to distinguish two frequencies in a time shorter than the one allowed by the Fourier limit is of direct relevance to signal processing. Additionally, external perturbations, such a magnetic or a gravity field, affect the system's quasienergy spectrum, making these perturbations detectable within the proposed scheme with superior sensitivity given the sub-Fourier features. This opens new avenues in magnetic field and gravity quantum sensing, with ultracold atoms in driven optical lattices representing a direct implementation  \cite{quantumsensing}.

The system considered here is a prototypical quantum ratchet consisting of a space-symmetric potential and a time-asymmetric oscillating force \cite{cubren16,reimann02}. Specifically, we consider a quasiperiodically driven quantum ratchet, the generalisation to quasiperiodic driving of the conventional periodic quantum ratchet \cite{denisov07,salger2009,denisov15,cubren16}. Specifically, we consider a quantum particle of mass $m$, subject to the following periodic potential 
\begin{equation}
V(x)=V_0\cos(2\pi x/L),
\label{eq:pot}
\end{equation}
and the following biharmonic driving force
\begin{equation}
F(t)=F_1\cos(\omega_1 t)+F_2\cos(\omega_2 t+\theta),
\label{eq:biharmF}
\end{equation}
where $\omega_1$ and $\omega_2$ are the driving frequencies, and $\theta$ the driving phase. In the simulations, reduced units are defined such that  $m=2\pi/L=5\hbar=1$.  For such a small value of the Planck constant, it is possible to associate Floquet states with invariant manifolds in phase space in the classical limit \cite{denisov07}.

For $F_{1,2}\neq 0$, $\omega_2\ne\omega_1$, and $\theta\neq n\pi$, with $n$ integer, all the relevant spatiotemporal symmetries of the system are broken \cite{cubren16,reimann02}, and the action of the external driving will set the particle into directed motion  \cite{denisov07,salger2009,denisov15,cubren16}. Quantitatively, the response of the system to the external driving will be quantified by the particle's average velocity.

\begin{figure}
\includegraphics[width=8cm]{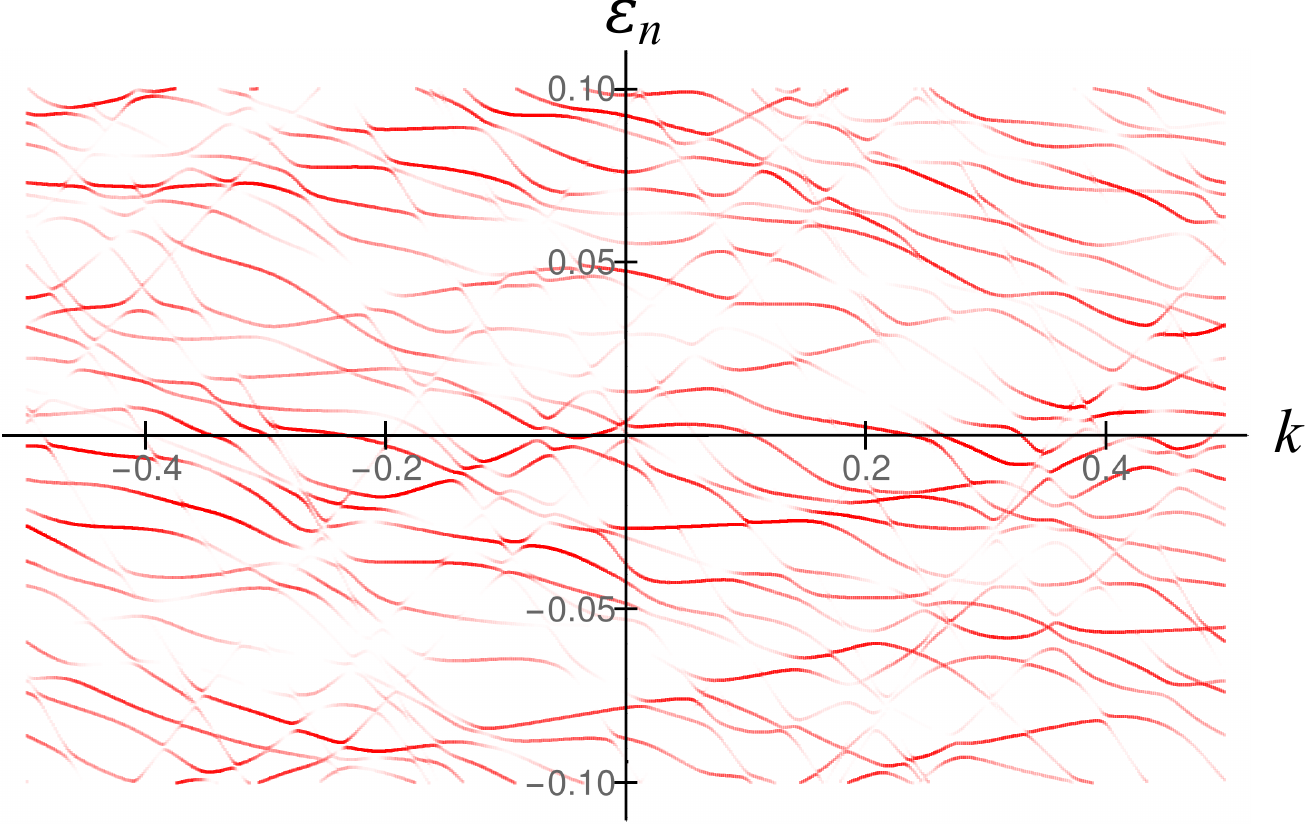}
\caption{Quasienergies $\epsilon_n$ as a function of the wave number $k$ for a spatially symmetric, as per Eq. (\ref{eq:pot}), quantum ratchet subject to the biharmonic force (\ref{eq:biharmF}), with $V_0=F_1=F_2=\omega_1=1$, $\omega_2=2\omega_1$, and $\theta=-\pi/2$. Each point has an opacity proportional to the absolute square of the projection of the Floquet-Bloch state $\psi_{k,n}(x,t)$ at $t=0$ onto the corresponding $\hbar k$-momentum eigenstate.
\label{fig:quasienergies}}
\end{figure}

Let us consider first the case of periodic driving, $\omega_2=\omega_1 p/q$, with $p$ and $q$ integer coprimes, and, thus, with a driving period $T=2\pi q/\omega_1$. Given the periodicity in time and space, we will adopt the Floquet-Bloch formalism \cite{hanggibook, cubero2018}. The quasienergy spectrum of the associated Floquet-Bloch states $\psi_{k,n}(x,t)$ --- where $\hbar k $ is the quasimomentum and $n$ a quantum number indexing the Floquet states --- is very rich, as shown in Fig.~\ref{fig:quasienergies} for a typical example, where the value of the driving phase $\theta$ was chosen to maximize the directed motion. Since the total number of quasienergy levels is infinite, in order to have a reasonable representation of them, depicting only the most relevant ones for the present discussion, the opacity of each point in  Fig.~\ref{fig:quasienergies} has been taken proportional to the absolute square of the projection of the Floquet-Bloch state $\psi_{k,n}(x,t)$ at $t=0$ onto the momentum eigenstate  $\exp(kxi)/\sqrt{L}$. In this sense, the most relevant Floquet-Bloch states are those with a large overlap with the corresponding momentum eigenstate. These are the states that are more readily prepared in a typical experiment.

The average current associated with a given Floquet-Bloch state $\psi_{k,n}$ is given by
\begin{equation}
v_n(k,\theta)=\frac{1}{T}\int_{t_0}^{t_0+T}\!\!\!\!\!\!\!\! dt\int_0^L \!\!\!\!\!dx\, \psi_{k,n}^*(x,t)\left(\!\!-\frac{\hbar}{im}\frac{\partial}{\partial x}\right)\psi_{k,n}(x,t).
\label{eq:momentum_ref}
\end{equation}
It can also be calculated from the slope in a plot of quasienergy vs. quasimomentum  via the relationship
\begin{equation}
v_n(k)=\frac{1}{\hbar}\frac{\partial \epsilon_{n}(k)}{\partial k}. \label{eq:momentum}
\end{equation}
The quasienergy spectrum of Fig.~\ref{fig:quasienergies} thus evidences  that the darker lines  correspond mainly to the lowest currents, with higher current values associated to states more difficult to observe (lower opacity  lines), as showing a small overlap with a momentum eigenstate.

\begin{figure}[t!]
\includegraphics[width=8cm]{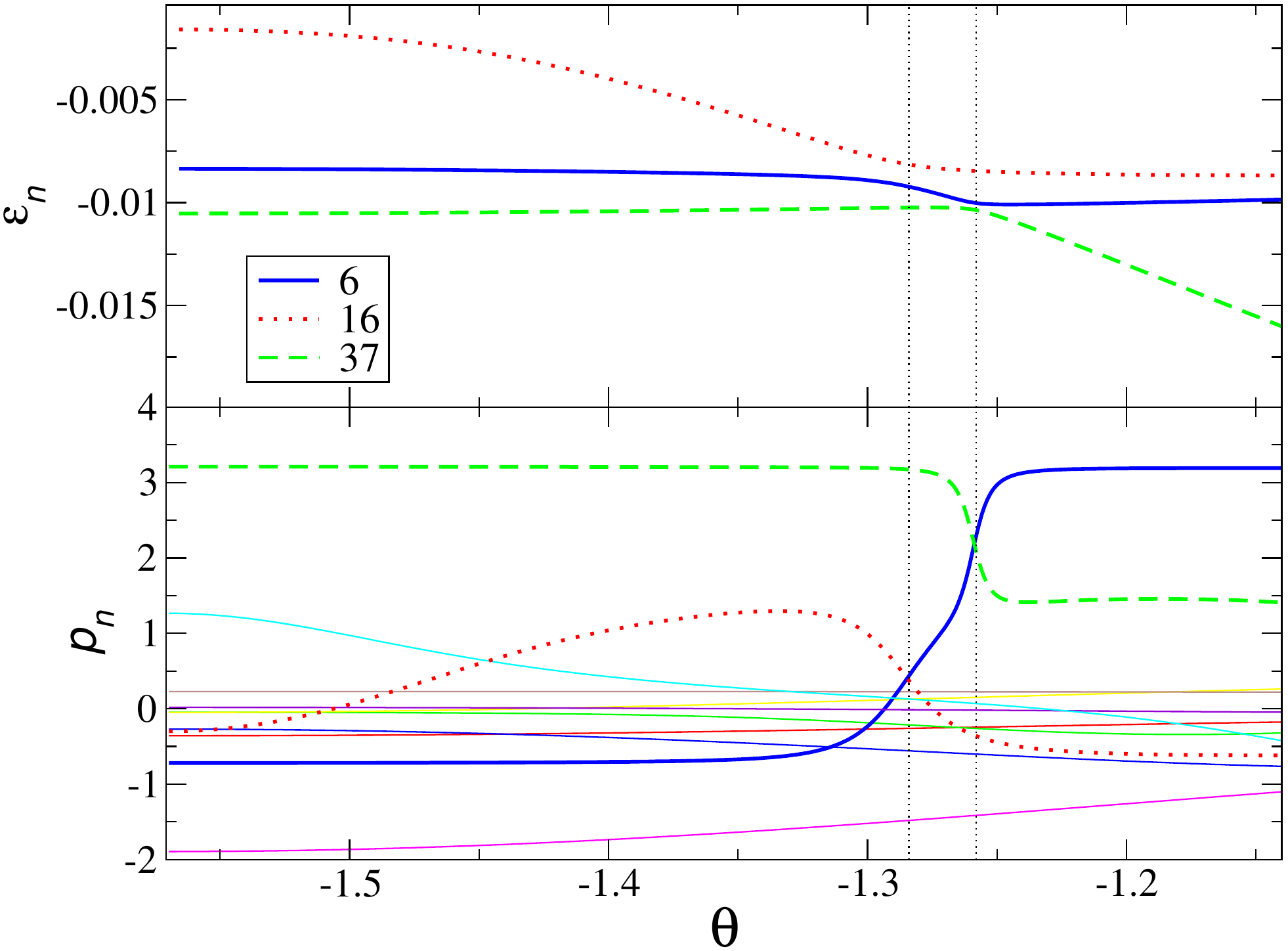}
\caption{Quasienergy and momentum $p_n=mv_n$ of selected Floquet-Bloch states as a function of the driving phase $\theta$ for the same system as in Fig.~\ref{fig:quasienergies} with a quasimomentum given by $\hbar k(\theta)=F_2(\sin\theta+1)/\omega_2$  --- which is obtained from (\ref{eq:kt}), see Supplemental Material \cite{suppl}.  These Floquet-Bloch states are the relevant ones for the asymptotic sub-Fourier narrowing mechanism discussed in the following.
The panels show two avoided crossings  ---indicated by vertical dotted lines--- between three levels: level 6 and 16, which has an appreciable component of the initial quantum state, and level 37, initially irrelevant. These two crossings provoke a fast change of identity of level 6, resulting in the current's rapid drop observed in Fig.~\ref{fig:startfromzero} at about $|\Delta \theta|=|\omega_2-2\omega_1|T_s\approx 0.3$. Only the quasienergies of the states undergoing sharp avoided crossing of relevance to the occurrence of sub-Fourier features are labeled, with the label scheme indicated in the top panel applying to both panels.
\label{fig:startfromzero:Flo}}
\end{figure}

Note however that the quasienergies are densely packed into the first Brillouin zone, $-\hbar\omega/2<\epsilon_n<\hbar\omega/2$, with $\omega=2\pi/T$ and $T=2\pi q/\omega_1$, a fact which determines multiple crossings between them. 
These crossings, unless a symmetry rules otherwise, are not real crossings but avoided crossings, in which the two quasienergies ---which can get very near each other---  never have identical values. The parameter region where the two energies are near is associated with a strong interaction and mixing between the two Floquet states, with the two states ultimately exchanging their roles.  

Avoided crossings occur by varying a system parameter. Here we are interested in avoided crossings generated by a variation in the driving phase, as will be shown in the following that these avoided crossings determine sub-Fourier narrowing in the long-time limit. Accordingly,
Fig.~\ref{fig:startfromzero:Flo} shows two avoided crossings, indicated by vertical lines, produced by varying the driving phase. 

A key element for the discussion of sub-Fourier mechanism identified in the following is the association of sharp variation in the particle velocity to avoided crossings generated by a variation in the driving phase. To this purpose, the plot of the quasienergy of Fig.~\ref{fig:startfromzero:Flo} is complemented by (bottom panel)  the corresponding particle momentum. This evidences the above mentioned correspondence between a sharp variation in momentum and an avoided crossing. 

We note in passing that to obtain maximum numerically accuracy, the values for the momentum in Fig.~\ref{fig:startfromzero:Flo} were calculated directly from Eq.~(\ref{eq:momentum_ref}), i.e. by calculating the Floquet-Bloch states, diagonalizing the evolution operator in a driving period, and then time averaging. For each value of the driving phase $\theta$, only the Floquet-Bloch states with $\hbar k(\theta)=F_2(\sin\theta+1)/\omega_2$ were considered, as these are the states which determine the asymptotic sub-Fourier narrowing, as it will be shown in the following.

An avoided crossing can take place within a very small interval of the varying parameter, thus involving very abrupt changes.  We demonstrate here, using numerical experiments, how those abrupt changes can be exploited in a finite time application, by increasing considerably the sub-Fourier sensitivity of the quantum  system. 

 \begin{figure}[t]
\includegraphics[width=8cm]{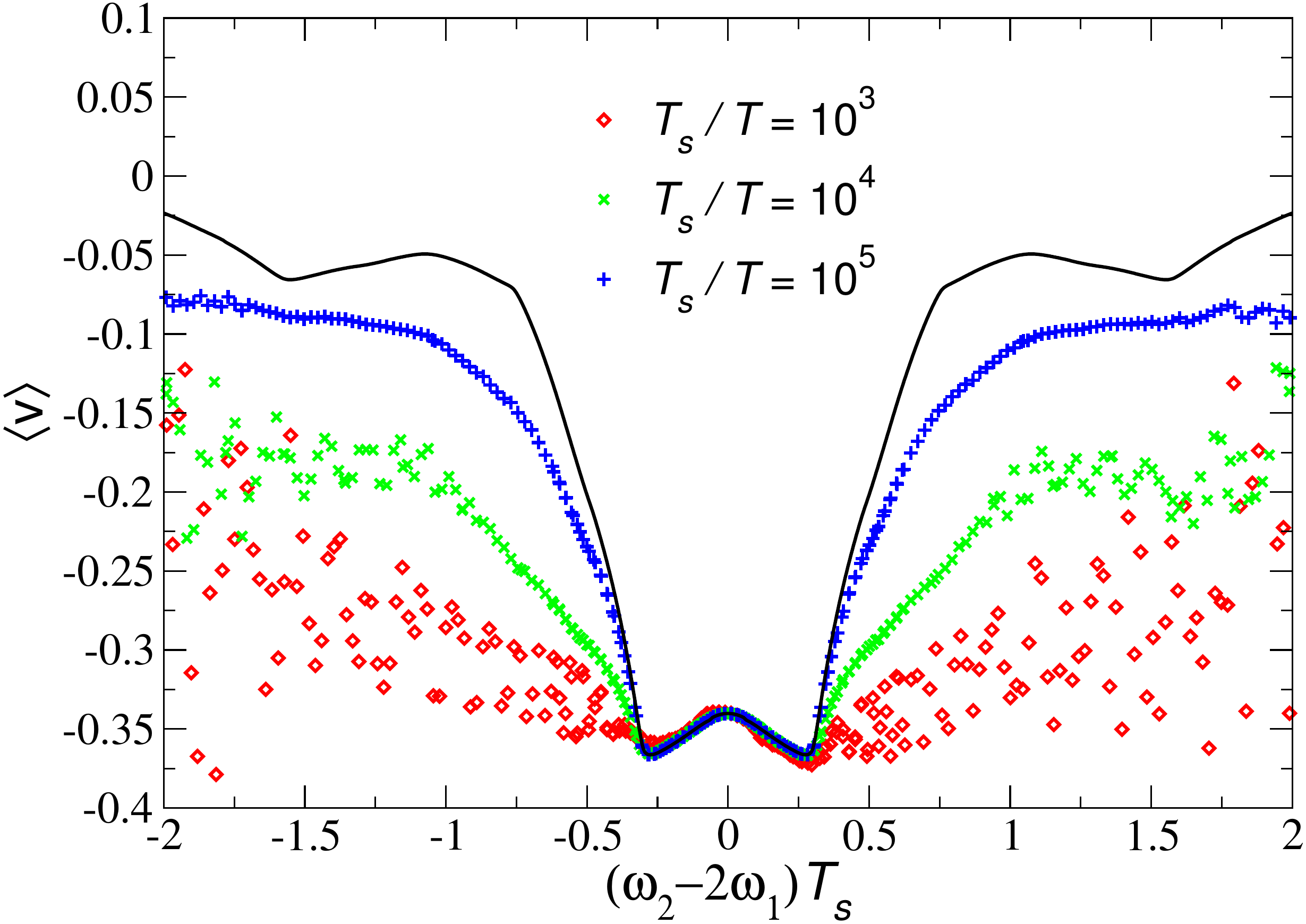}
\caption{Finite-time current as a function of the second driving frequency $\omega_2$ for 
a system with same parameters as in Figs.~\ref{fig:quasienergies}--\ref{fig:startfromzero:Flo} with $\theta=-\pi/2$,
%a spatially symmetric (\ref{eq:pot}) quantum ratchet subject to the biharmonic force (\ref{eq:biharmF}), with $V_0=F_1=F_2=\omega_1=1$, and $\theta=-\pi/2$, 
starting from a uniform wave function $\psi(x,0)=$constant. 
The solid line is the asymptotic prediction (\ref{eq:vfinite}), and the points correspond to several observation times. The plot shows an improvement about $10\times$ better than the Fourier width $\Delta\omega_2=2\pi/T_s$.
\label{fig:startfromzero}}
\end{figure}

In order to examine linewidth narrowing, and identify eventual sub-Fourier mechanisms,  it is necessary to determine the response of the system as a function of the frequency of the perturbation. To this purpose, we consider the quantum ratchet of Eqs.~(\ref{eq:pot})--(\ref{eq:biharmF}), but with a second driving frequency that is not necessarily in a rational ratio with the first one. The driving force can thus not be periodic at all. In any real experiment, the time spent measuring the current can not be taken to infinity, but to a ---supposedly large--- observation time $T_s$. We are interested in the dependency of the finite-time current $v_{T_s}=\frac{1}{T_s}\int_{t_0}^{t_0+T_s}\!\!dt \,v(t)$ ---where $v(t)=\langle \psi(t) | (p/m)| \psi(t)\rangle$ is the expected velocity at instant $t$---
as a function of the driving frequencies. 

It was shown in \cite{cubero2018} that the finite-time current can be expressed in terms of the current $v_n(k,\theta)$ of Floquet-Bloch states of nearby time periodic systems, 
\begin{equation}
v_{T_s}\sim \frac{1}{\delta \omega_2T_s}\int_{\theta_0}^{\theta_0+\delta \omega_2 T_s}\! d\widetilde{\theta}\, 
\sum_{k_0,n}|C_{k_0,n}|^2 \, v_{n}\!\!\left(k(\widetilde{\theta}),\widetilde{\theta}\right),
\label{eq:vfinite}
\end{equation}
where
\begin{equation}
k(\widetilde{\theta})=k_0+\lim_{\delta\omega_2\to0}\int_{t_0}^{t_0+T\lfloor\frac{\widetilde{\theta}-\theta}{\delta\omega_2T}\rfloor}\!\! dt^\prime F(t^\prime)/\hbar,
\label{eq:kt}
\end{equation}
 $\delta\omega_2=\omega_2-\omega_1 p/q$ is the frequency distance to the nearby periodic drive, with $p$ and $q$ coprime integers, and $\theta_0=\theta+\omega_2t_0$. Equation (\ref{eq:vfinite}) is valid in the asymptotic limit $T_s\to\infty$, $\delta\omega_2T_s$=constant.
The population $|C_{k_0,n}|^2=|\langle{\psi}_{k_0,n}(t_0)|\psi(t_0)\rangle|^2$  depends on the probability distribution of Floquet-Bloch states at initial time $t_0$. 

 We stress that the long-time resonant behaviour of $ v_{T_s}$ as a function of the driving frequency is determined by the Floquet-Bloch states selected by $k(\theta)$ of Eq.~(\ref{eq:kt}) as a function of the driving phase $\theta$, as per Eq.~(\ref{eq:vfinite}). 
%%
% {\sout {\cor This clarifies why in Fig. \ref{fig:startfromzero:Flo} we explored avoided crossings as a function of $\theta$ for specific Floquet-Bloch states indentified by  $\hbar 
% k(\theta)=F_2(\sin\theta+1)/\omega_2$. This is indeed the value of $k(\theta)$ determined by Eq. \ref{eq:kt} for the driving (\ref{eq:biharmF}) considered here.}} 
%  {\mycor XXX(I think the above is a bit too much, not really needed already after the above three added clarifying remarks)XXX} 

\begin{figure}[b!]
\includegraphics[width=8cm]{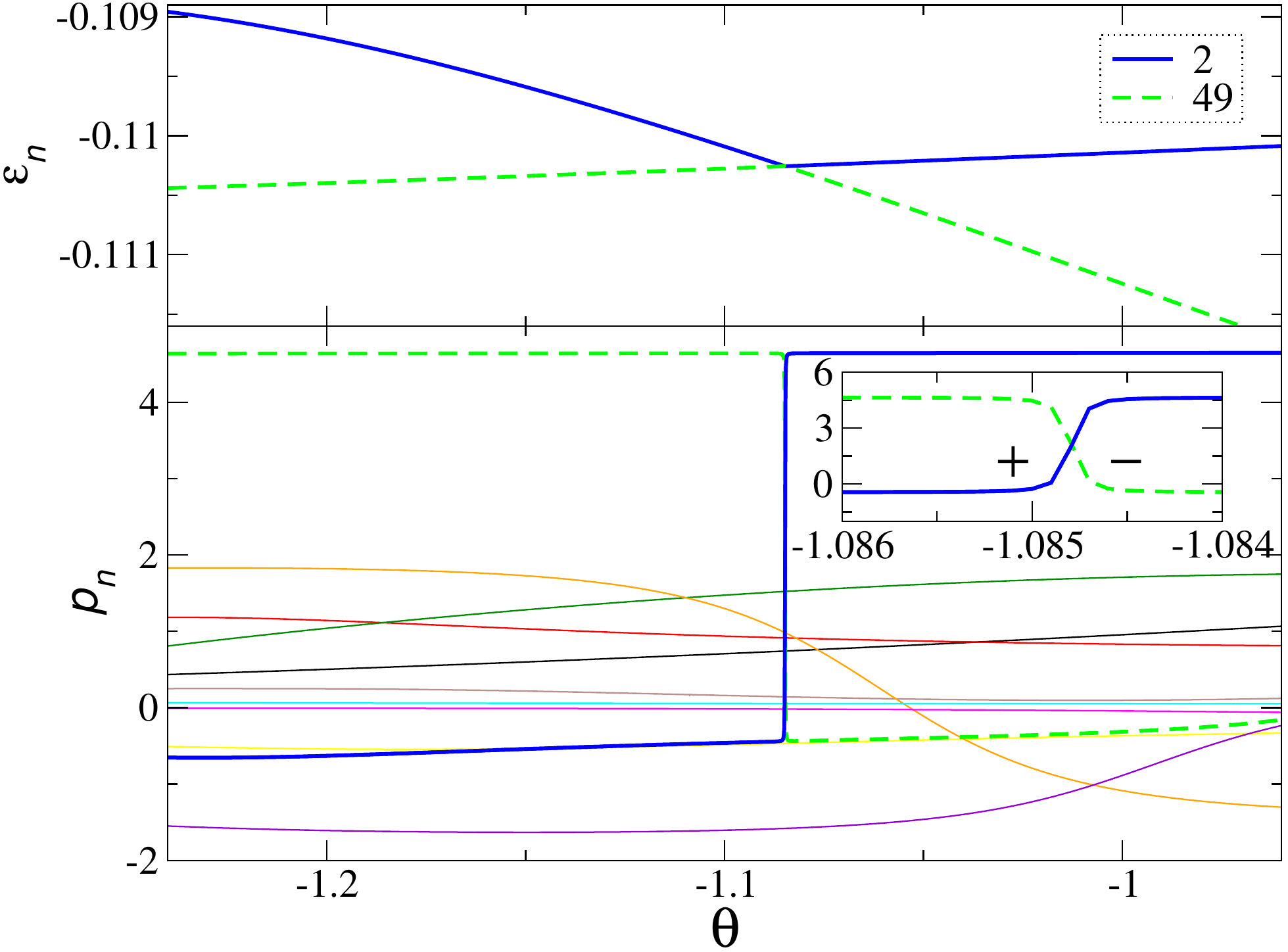}
\caption{A very sharp avoided crossing. Quasienergy (top) and current (bottom) of selected Floquet states as a function of the driving frequency $\theta$ for a system with $V_0=F_1=1$, $F_2=2.4$, $\omega_1=1.2$, $\omega_2=2\omega_1$, with $k(\theta)$ given by (\ref{eq:kt}), starting from $k_0=0.4236$ and $\theta= -1.247$. 
Only the quasienergies of the states undergoing sharp avoided crossing of relevance to the occurrence of sub-Fourier features are labeled, with the label scheme indicated in the top panel applying to both panels.
The inset shows an enlargement of the region around the avoided crossing, with the $+$ ($\theta=-1.0851$)
 and $-$ ($\theta=-1.0845$)
 indicating the driving phases and starting points in the simulations of Fig.~\ref{fig:startfromk}.
\label{fig:startfromk:Flo}}
\end{figure}

Figure~\ref{fig:startfromzero} shows the frequency dependency of the finite-time current for a system with the same parameters as discussed in Figs.~\ref{fig:quasienergies} and \ref{fig:startfromzero:Flo}, starting from a uniform wave function $\psi(x,0)$=constant ---i.e. $k_0=0$. A very good agreement is found between the numerically calculated finite-time current of the ratchet and the asymptotic prediction (\ref{eq:vfinite}), based on Floquet-Bloch states of nearby periodic systems, in the expected limit, that is, for small $\delta\omega_2$ and large $T_s$. 

Note the width of the resonance in Fig.~\ref{fig:startfromzero} is smaller, by a factor of about 10, over the expected Fourier width $\Delta\omega_2T_s=(\omega_2-2\omega_1)T_s=2\pi$. A classical ratchet is also expected to be sub-Fourier \cite{cubero2013,cubero2014}, more specifically $\Delta\omega_2T_s=2\pi/q$. However, in the example considered here $q=1$ and the classical result agrees with the Fourier width---being in fact what is observed in the simulations of \cite{cubero2012}. Thus, there is evidence of sub-Fourier narrowing determined by the quantum nature of the system.

 The sub-Fourier behavior of  Fig.~\ref{fig:startfromzero} can be explained in terms of avoided crossings between Floquet states. The bottom panel of Fig.~\ref{fig:startfromzero:Flo} shows the average momentum $p_n=mv_n$ of the first 10 Floquet states, together with the 16th and 37th. The states have been sorted according to their initial probability $|C_n|^2$. The top panel focuses on the quasienergy of the three levels involved in two avoided crossings, which are indicated by vertical dotted lines. Of the three,  the first two levels have an apreciable probability $|C_6|^2=0.063$ and $|C_{16}|^2=0.019$, whereas the third is negligible  $|C_{37}|^2=4\cdot10^{-7}$. %$|C_{16}|^2=0.01878$, $|C_{37}|^2=0.0000003645$
However, the two avoided crossings at about $\theta\sim-1.27$ provoke a change of identity in the state 6, first swapping with 16, and ending up with the quasienergy and current of the level 37. The current of the state 6 thus changes drastically in a small interval, being responsible of the abrupt drop observed in Fig.~\ref{fig:startfromzero}. 

Thus, the sub-Fourier narrowing is determined by sharp avoided crossings as a function of the driving phase. The emergence of this mechanism for sub-Fourier narrowing in a driven quantum system is the central result of this work.

The quasienergy spectrum is sufficiently rich to allow for much more sensitive sub-Fourier behavior than the one presented above. Figure~\ref{fig:startfromk:Flo} shows an example of an extremely  sharp avoided crossing between a relevant Floquet state, level 2, with an (initially) irrelevant one, level 49, which carries a considerably larger current.

\begin{figure}
\includegraphics[width=8cm]{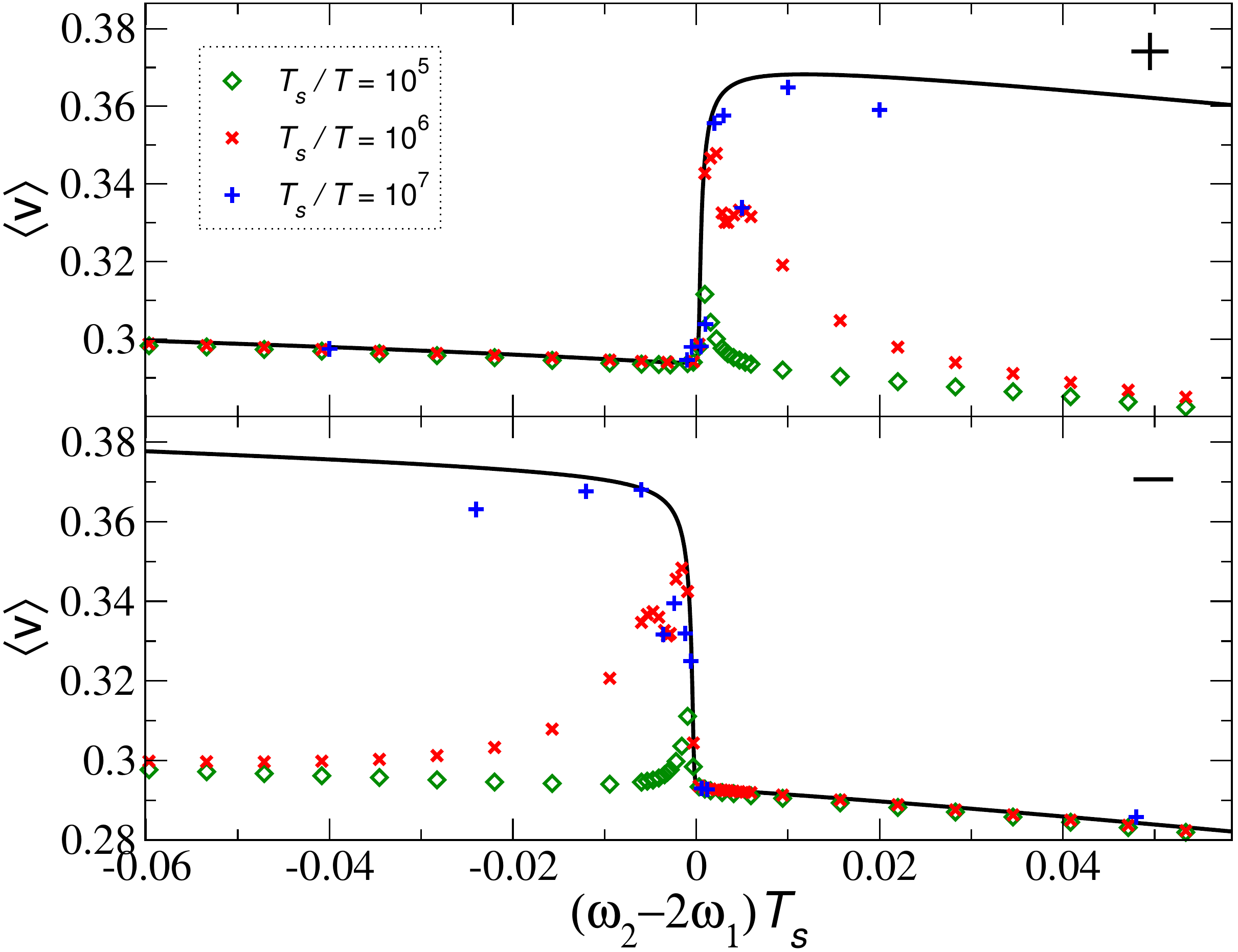}
\caption{Finite-time current exhibiting a frequency sensitivity about $1000\times$ finer than Fourier resolution. The biharmonic drive has the same parameters as in Fig.~\ref{fig:startfromk:Flo}. 
Two sets of simulations are shown: $+$ (top) with driving phase $\theta=-1.0851$, and $-$ (bottom) with $\theta=-1.0845$, both starting at $t=0$ from the corresponding wave numbers implied in  Fig.~\ref{fig:startfromk:Flo}.    %$k= 0.742012$ for $+$ and $k=0.743413$ for $-$
Before the application of the biharmonic driving, a constant force is applied during $2T$ in order to start at $t=0$ with the desired wave number. 
The solid lines are the asymptotic predictions (\ref{eq:vfinite}), and the points correspond to several  different observation times. 
\label{fig:startfromk}}
\end{figure}

In order to take advantange of the avoided crossing of Fig.~\ref{fig:startfromk:Flo} in a finite-time measurement, we can make use of a convenient result, which states that if the system $\psi(x,t_0)$ is a Bloch state with wave number $k_0$, then $\psi(x,t)$ is also a Bloch state with wave number $k(t)=k_0  +\int_{t_0}^t\!\!dt^\prime\,F(t^\prime)/\hbar$. Therefore, starting from a uniform wave function $\psi(x,t_0)$=constant ($k_0=0$), we can obtain a Bloch state with a desired target wave number $k$ by applying a constant force $F_0=\hbar k/\Delta t$. 

Figure~\ref{fig:startfromk} shows the result of such approach.  All simulations start from a uniform wave function ($k=0$) at $t=-2T$, are moved to the desired $k$-point %($k= 0.742012$ and $k=0.743413$ for the points denoted by $+$ and $-$ in  Fig.~\ref{fig:startfromk:Flo}, respectively) 
 by application of a constant force during a time interval $2T$, and then ($t=0$) the biharmonic driving (\ref{eq:biharmF}) is switched on for a time $T_s$ to measure the ratchet current. 
The measures starting from the point $+$ in  Fig.~\ref{fig:startfromk:Flo} find almost no current variation for values $\delta\omega_2<0$, while positive $\delta\omega_2$ are detected with great sensitivity due to the avoided crossing. A similar behavior is found  for the detection of negative $\delta\omega_2$ in the simulations shown in the bottom panel of Fig.~\ref{fig:startfromk}. Compared to Fourier resolution, the quantum ratchet demonstrates here a sensitivity about $1000$ times finer, a result that is intimately related to the coherent nature of the quantum dynamics.

In conclusion, this work examined the long-time response of a driven quantum system. Sub-Fourier behaviour was identified, with the response of the system, examined as a function of the perturbation frequency, narrowing faster for increasing interaction time than the Fourier limit. The occurrence of sub-Fourier narrowing was traced back to a quantum-mechanical mechanism, with the sub-Fourier dynamics associated to sharp avoided crossing observed as a function of the driving phase. Inspection of the quasi-energy spectrum thus allows for the identification of sub-Fourier features in the system's response. Besides its  fundamental interest, the mechanism identified here may find application in metrology and sensing, where the width of the relevant resonance determines the utimate performance of the measurement.

%
%
%\\
%\\
%{\bf Supplementary information} is available in the online version of the paper
%\\
%\\
%{\bf acknowledgments} 
\begin{acknowledgments}
Financial support from  the Ministerio de Econom\'ia y Competitividad of Spain, Grant No. FIS2016-80244-P is acknowledged. 
\end{acknowledgments}
%\\
%\\
%{\bf Author information} Reprints and permission information is available at www.nature.com/reprints. The authors declare no competing financial interests. Readers are welcome to comment on the online version of the paper. Correspondence and requests of materials should be addressed to D.C. (dcubero@us.es) or F.R. (f.renzoni@ucl.ac.uk). 

%%%%%%%%%%%%%%%%%%%%%%%%%%%%%%%%%%%%%%%%%%%%%%%%%%%%%%%%%%%%%%%%%%%%%%%%
 
%%%%%%%%%%%%%%%%%%%%%%%%%%%%%%%%%%%%%%%%%%%%%%%%%%%%%%%%%%%%%%%%%%%%%%%%

%\bibliography{book_references}

%\onecolumngrid

\end{document}